# Theoretical Study of the Hard Photons Flow Emission Rate At $s\bar{b} \rightarrow \gamma g$ Systems in Annihilation Processes in One Loop

Hadi J. M. Al-Agealy[1], Maher Sami Saleh[2]

[1]Department of Physics, College of Education for pure Science, Baghdad University, Baghdad, Iraq
[2] Ministry of Education, Baghdad, Iraq

## Abstract

In this paper, we evaluate and analyze hard photons flow emission at $s\bar{b} \rightarrow \gamma g$ quarks system interaction in annihilation processes due to a quantum field theory in high energy physics. Two quantum wave vectors are assumed to localize quark and anti-quark in quantum space for higher thermal energy system. Quantum postulate theory and perturbative theory take into account the photonic flow emission rate that gets produced according to quantum model. The photonic flow rate is evaluated depending on physical parameters: activation coupling strength, electric charge of quark system, fugacity, transition momentum, quantum color number, photonic energies, total charge quark system, Euler factor, critical temperature and temperature of the quarks system. The photonic flow rate for $s\bar{b} \rightarrow \gamma g$ quarks system production in one loop Annihilation processes depends on the activation coupling strength, thermal energy, critical temperature, and fugacity. The rate increases with decreases in activation coupling strength and the critical temperature and with increases in the temperature and vice versa. However, the photonic flow rate increases with the fugacity at one-loop based on Juttner distribution and vice versa.

**Keywords**: Flow emission rate, Annihilation Processes

# Introduction

In a few years ago, the research in elementary particle increases vertically and horizontally, experimentally and theoretically. The ultra-relativistic heavy ion collision is used to have a collision process for two heavy ions. The collision is head-on and goes at high relativistic speed [1].

However, the physics at different places around the world have been researching in elementary particles, with a view to discussing the constituents of proton and neutron matter [2]. Furthermore, advances in science and technology give us much more information about the formation and the development of matter. They experimentally prove that the elementary particle was regarded as never been a breakable constituent or never consists of any smaller particle [3].

In 1964, Gell-Mann and Zweig introduced the quarks model to investigate the matter of hadrons [4]. According to quark model, the quarks assume to be bound with each other in the hadronic material at lower densities and temperatures, the nucleons could be melting in the collision under huge degrees of pressure and temperature. The state of quark-gluon matter is forming in heavy ion collisions and that could be investigated at strong interaction [5]. The standard model theory has field quantum theory, it describes the interaction between quarks and gluons due to exchange the bosonic gauge field. It builds up by the quantum chromodynamics theory QCD and electroweak theory [6]. According to the standard model theory, the hadrons consist of fundamental blocks named quarks; they exist and are bound in a quantum state. The standard model indicates that there are six quarks with increasing masses. On the other hand, there are six anti-quarks. The quarks and anti-quarks are equivalent in mass and opposite in charge [7]. At the time, the scientists knew that protons and neutrons are built by fundamental basic particles named quarks in hadrons. The Hadron is subdivided to baryons groups that contain three quarks and mesons groups contain pairs of quark-antiquark particles.

In both cases, the quarks or antiquarks in two baryons groups and mesons groups are confined due to gluons [8]. The Quantum Chromodynamics field theory is a good theory that introduced discussion for the interaction of quarks and gluons. Generally, it's assumed that fundamental particles defined due to localized quantum fields. Specifically, the physical terms were in consensus to define the local fields using quantum field theory and quantum chromodynamics theory QCD. The QCD is the theory that has much more discussed strong nuclear interactions depending on quarks behavior [9]. According to the field theory, we classify the quarks into three generations for six quarks such that Up and Down in the first generation, Charm and Strange in the second generation and Bottom and Top in the third generation. Quarks have a spin that equals to $1/2\ \hbar$, but the mediate gauge of the quarks called gluons has $1\ \hbar$ spin. Quarks and leptons are one family named fermions, while the gluons are called bosons [10].

**Theory**

Under the quantum chromodynamics theory, the flow of photons emission due to quarks interaction can be studied and investigated according to the quantum postulate and perturbative theory. The quark and anti-quark are assumed to be localized state vector in spacetime state. The flow photons emission rate can be given as [11].

$$\epsilon \frac{dn_\gamma}{d^4x d^3P} = -\frac{1}{(2\pi)^3} F^B(\epsilon,T) \text{Im} \prod_{\mathcal{MN}}^{\mathcal{T}}(\epsilon, \lambda_{q,\dot{q}}, T) \ldots\ldots\ldots\ldots\ldots (1)$$

Where $F^B$ is the Bosonic of the distribution function and, $\text{Im} \prod_{\mathcal{MN}}^{\mathcal{T}}(\epsilon, \lambda_{q,\dot{q}}, T)$ is the photon self-energy propagation, then Bosonic distribution function will be reformulated as [12].

$$F^B(\epsilon,T) = \frac{1}{(e^{E_\gamma/T}-1)} \ldots\ldots\ldots\ldots (2)$$

Where $E_\gamma$ is the energy of flow photons, T is the temperature of quarks system. Furthermore, the self-energy propagations have been represented using the spectral function. It could be estimated using [13].

$$\text{Im}\,\Pi^{\mathcal{T}}_{\mathcal{MN}}(\epsilon, \lambda_{q,\acute{q}}, T) = 2q^2 \sum e^2_{QCD} T \times \frac{1}{(2\pi)^3} \int Tr[d^*(\kappa)\Lambda^\mu(\kappa, \overline{\kappa}, -p).d(\kappa - P)\Lambda^{\mathcal{T}}(-\overline{\kappa}, -\kappa, p)]d^3\kappa` \ldots\ldots\ldots (3)$$

Where $q$ is the constant of the electromagnetic field, $e_{QCD}$ is the quarks electric charge, T is the thermal energy of the system, $\Lambda$ is the density of photons state, and $d$ is the propagation functions. The propagation of the quarks is given by [14].

$$d(\kappa - P) = \frac{1}{2\pi}\int_{-\infty}^{\infty} \frac{D(\omega, \overline{\kappa})}{\omega - p - i\kappa_o} d\omega \ldots\ldots(4)$$

And

$$d^*\kappa = \frac{1}{2\pi}\int_{-\infty}^{\infty} \frac{D^*(\omega, \overline{\kappa})}{\omega - i\kappa_o} d\omega \ldots\ldots(5)$$

Inserting Eq. (4) and Eq. (5) in Eq. (3) to results

$$\text{Im}\,\Pi^{\mathcal{T}}_{\mathcal{MN}}(\epsilon, \lambda_{q,\acute{q}}, T) =$$

$$-\frac{10\pi}{3} q^2 \sum e^2_{QCD} (e^{E_\gamma/T} - 1) \times \frac{1}{(2\pi)^3} \int d^3\kappa \int_{-\infty}^{\infty} d\omega \int_{-\infty}^{\infty} d\tilde{\omega}\, \delta(\epsilon - \omega - \tilde{\omega})[F_{(q)}(\omega).F_{(\acute{q})}(\tilde{\omega})]Tr[\mathfrak{D}^\mu(\kappa,\overline{\kappa},-p)D^*(\omega,\overline{\kappa})D(\omega-\epsilon,\overline{\kappa}-\overline{p})\mathfrak{D}^{\vartheta}(-\overline{\kappa},-\kappa,p)]$$

$$\ldots\ldots\ldots (6)$$

Where $F_{(q)}(w)$ and $F_{(\acute{q})}(\tilde{w})$ are the distribution Jouttner functions may be written as [15].

$$F_{(q)} = \frac{\lambda_q}{e^{\frac{E_1}{T}}+1}, \quad and \quad F_{(\acute{q})} = \frac{\lambda_{\acute{q}}}{e^{\frac{E_2}{T}}-1} \ldots\ldots\ldots (7)$$

Then the Eq. (6) will be rewritten using Eq. (7) to became

$\text{Im} \, \Pi^T_{MN}(\epsilon, \lambda_{q,\dot{q}}, T) =$

$-\frac{10\pi}{3} q^2 \sum e^2_{QCD} \left(e^{E_\gamma/T} - 1\right) \times$

$\int \frac{d^3\kappa}{(2\pi)^3} \int_{-\infty}^{\infty} \frac{d\omega}{2\pi} \int_{-\infty}^{\infty} \frac{d\tilde{\omega}}{2\pi} \delta(\epsilon - \omega - \tilde{\omega}) \frac{[F_{(q)}.F_{(\dot{q})}]}{\omega+\tilde{\omega}-i} Tr[\mathfrak{D}^\mu(\kappa, \overline{\kappa}, -p) D^*(\omega, \overline{\kappa}) D(\omega - \epsilon, \overline{\kappa} - \overline{p}) \mathfrak{D}^\vartheta(-\overline{\kappa}, -\kappa, p)] \dots \dots \dots (8)$

The Eq. (8) can be simplified using higher mathematical physics to results [14].

$\text{Im} \, \Pi^T_{MN}(\epsilon, \lambda_{q,\dot{q}}, T) =$

$-4\pi \frac{10\pi}{3} q^2 \sum e^2_{QCD} \left(e^{E_\gamma/T} - 1\right) \times \int \frac{d^3\kappa}{(2\pi)^3} \int_{-\infty}^{\infty} \frac{d\omega}{2\pi} [F_{(q)}.F_{(\dot{q})}] \times \frac{1}{2\kappa} \delta(\cos\theta - \frac{\omega}{\kappa}) [D^*_+(\omega, \overline{\kappa})(-1 + \frac{\omega}{\kappa}) + D^*_-(\omega, \overline{\kappa})(-1 - \frac{\omega}{\kappa}) \dots \dots \dots (9)$

However, we must be using the condition, $E_q + E_{\dot{q}} = E_\gamma$ and $E_\gamma \gg T$ for large energy. Under this condition, we use the approximation to replace.

$F_{(q)}.F_{(\dot{q})} \rightarrow \frac{\lambda_q}{e^{\frac{E_q}{T}}+1} \cdot \frac{\lambda_{\dot{q}}}{e^{\frac{E_{\dot{q}}}{T}}-1} \approx \lambda_q \lambda_{\dot{q}} e^{\frac{E_q}{T}} e^{\frac{E_{\dot{q}}}{T}} \approx \lambda_q \lambda_{\dot{q}} e^{-\frac{E_\gamma}{T}} \dots \dots \dots (10)$

Then we can simplify the Eq. (9) using Eq. (10) to

$\text{Im} \, \Pi^T_{MN}(\epsilon, \lambda_{q,\dot{q}}, T) = -4\pi \frac{5}{12\pi^2} q^2 \sum e^2_{QCD} \left(e^{E_\gamma/T} - 1\right) \times \left[\lambda_q \lambda_{\dot{q}} e^{-\frac{E_\gamma}{T}}\right] \{\int \kappa d\kappa \times [\sigma_+(\kappa)\left(-1 + \frac{\omega}{\kappa}\right) + \sigma_-(\kappa)\left(-1 - \frac{\omega}{\kappa}\right) + \int \kappa d\kappa \beta_-(\omega, \kappa) \times \theta(\kappa^2 - \omega^2)\} \dots \dots \dots (11)$

Interestingly, the final term in integral (11) refers to the correlation factor and can be solved by assuming $\theta = \frac{2}{\pi} \frac{Q_0(\sinh\eta) - Q_1(\sinh\eta)}{1-\tanh\eta}$ and $y_c = \frac{2}{\pi} \frac{k_c^2}{m_q^2}$ and we solve to [16].

$\mathbb{C}_{cor} = \int \kappa d\kappa \beta_\pm(\omega, \kappa) \times \theta(\kappa^2 - \omega^2) \cong m_q^2(-1 - C_E) \dots \dots \dots (12)$

Where $C_E = 0.577216$ [17].

However, we can use equality formula to results [18].

$$\int [\sigma_+(\kappa)\left(-1 + \frac{\omega}{\kappa}\right) + \sigma_-(\kappa)\left(-1 - \frac{\omega}{\kappa}\right)]\kappa d\kappa =$$

$$2m_q^2 \int_\kappa^\mu \frac{\omega_+(\kappa) - \omega_-(\kappa)}{m_q^2} d\kappa = Ln\frac{\mu^2}{\kappa^2} \quad\quad\quad (13)$$

Inserting the Eq. (13) and Eq. (12) in Eq. (11) to

$$Im\ \Pi_{MN}^T(\epsilon, \lambda_{q,\dot{q}}, T) = -4\pi \frac{5}{12\pi^2} q^2 \sum e_{QCD}^2 \left(e^{E_\gamma/T} - 1\right) \times \left[\lambda_q \lambda_{\dot{q}} e^{-\frac{E_\gamma}{T}}\right] [m_q^2 Ln\frac{\mu^2}{k^2} +$$

$$m_q^2(-1 - C_E)] \quad\quad\quad\quad (14)$$

Substituting Eq. (14) in Eq. (1) to gate current photons rate equation.

$$\mathfrak{F}_{ANN}(\alpha_{ASC}, E_\gamma, T) = \frac{1}{8\pi^4} \times \frac{5}{3} q^2 \sum e_{QCD}^2 \times \left[\lambda_q \lambda_{\dot{q}} e^{-\frac{E_\gamma}{T}}\right] \times m_q^2 [Ln\frac{\mu^2}{k^2} - 1 -$$

$$C_E] \quad\quad\quad\quad (15)$$

Where $\alpha = q^2/4\pi$ and $m_q^2 = \frac{g^2 C_F T^2}{4}$ [19].

And the activation nuclear strong constant $\alpha_{ASC} = g^2/4\pi$ [20]. For the limit of $k \sim gT$, and $\mu \sim \sqrt{2ET}$ the Eq. (12) becomes.

$$\mathfrak{F}_{ANN}(\alpha_{ASC}, E_\gamma, T) = \frac{30\alpha_{em}\alpha_{ASC}}{27\pi^2} \sum e_{QCD}^2\ T^2 \lambda_q \lambda_{\dot{q}} e^{-\frac{E_\gamma}{T}} [Ln\frac{2E}{4\pi\alpha_{ASC}T} - 1 - C_E]\ldots$$

(16)

The activity strength coupling $\alpha_{ASC}$ for quarks interaction will be [21].

$$\alpha_{ASC} = \frac{6\pi}{(33-2N_f)ln(\frac{P_{AC}}{T_c})} \quad\quad\quad\quad (17)$$

Here $P_{AC}$ is the momentum transition, $T_c$ is critical phase temperature, and $N_f$ is the quantum flavor number.

**Results**

Annihilation of quark anti-quark interaction is the basic and fundamental photons producing processes. A theoretical investigation treatment based on quantum chromodynamics theory adapted to calculation and studies of the behavior of quarks at annihilation processes and analysis of the rate of photons. Based on the quantum theory, the photons flow rate has been evaluated as a function of the activation coupling strength $\alpha_{ASC}$, electro dynamic constant $\alpha_{em}$, photons energy $E_\gamma$, the electric charge of quarks $e_{QCD}$, flavour number $N_F$, temperature of system T. The photons rate and all coefficients are evaluated using MATLAB program. Firstly, we evaluate the electric charge of the quark system. With such combination of quark anti-quark interaction, it is possible to estimate the effective charge of quarks system by using the summation of square charge $\sum e_{QCD}^2$ for a $s\bar{b} \rightarrow \gamma g$ system, and results is $\sum e_{QCD}^2 = \frac{2}{9}$. More details of properties and behavior of the quarks in strong interaction has been depending on the activation coupling strength constant $\alpha_{ASC}$, we have been able to do the calculation using Eq. (17). based on total flavours quarks number of systems to be 8 and critical temperature $T_c = 155$MeV, and 185MeV and the photons energy in limited range scale $4 \leq E_\gamma \leq 8\ GeV$, results of $\alpha_{ASC}$ has listed in the table (1) for critical temperature 155MeV, and 185MeV for $s\bar{b} \rightarrow \gamma g$ quarks systems.

Annihilation process for quark anti-quark interaction gives us important information for photons production that is evaluated according to Eq. (16) for $s\bar{b} \rightarrow \gamma g$ systems using MATLAB program. The flow photons emission for $s\bar{b} \rightarrow \gamma g$ interaction at annihilation processes uses the photons energy ranging $E_\gamma = 4 \rightarrow 8$ GeV [22]. and uses values for the activation coupling strength $\alpha_{esc}$ from table (1), thermal energy $T = 150$MeV, $T = 175$MeV, $T=200$MeV, $T=225$MeV and T=250MeV, taken the critical temperature $T_c$= 155MeV and 185MeV and fugacity parameters $\lambda_q = 0.068$, and $\lambda_{\bar{q}} = 0.02$ .Inserting all these parameters in Eq.(16) and solving them by MATLAB program give the results listed in tables (2) ,(3) ,(4) and (5) for $s\bar{b} \to \gamma g$ quarks systems.

**Discussion**

The flow photons emission rate in hard quark anti-quark interaction of annihilation processes at quantum media has been investigated and evaluated theoretically. In the process, we investigate and study the photonic emission at $s\bar{b} \to \gamma g$ quarks system at annihilation processes theoretically. It will be an evaluation of the photon emission rate in a high energy quarks medium under space-time of thermal system. It has been studied based on color quantum theory. The yield of photons production rate at quark-antiquark annihilation processes is according to energetic equilibrium. It is the lowest contribution that's corresponding to a low-lying loop and the quark propagator was taken into account. On the other hand, the structure of strong forces at high energy can be understood depending on its interaction; the quarks physic introduced much more information to study the constituents of matter. As shown in the tables (2), (3), (4) and (5), the photonic rate has been generated by the many parameters which are in the Eq. (16) and in the activation strong coupling in Eq. (17). The flow photons rate spectrum is obtained using the flavor number, color charge, electric charge, photons energy, thermal energy, and fugacity over space-time of quarks system by adopting the quantum chromodynamics theory with perturbative quark behavior due to activation coupling strength for the different nuclear system. The evaluation of photons rate from annihilation processes at a fixed phase has been determined by applying the Juttner distribution for quarks and anti-quarks in quantum space. Quantum chromodynamics theory is investigated by the strong nuclear force at higher energies for the quarks interaction system at annihilation processes by assuming that the two quarks are described by wave vectors in color quantum space.

In contrast, the photons yield rate is emission for the collision of the quark anti-quark under the color field of annihilation processes; it has been an evaluation depending on the rate expression in Eq. (16). As would be shown with the analysis of photonic rate expression at annihilation processes, it was enhanced exponentially $e^{-\frac{E_\gamma}{T}}$ with photons energy $E_\gamma$ and temperature of the system $\frac{1}{T}$. At fixed critical temperature, the photons rate production increase with the decreasing of the energy of the photon $E_\gamma$ and temperature of system T.

The critical temperature of the system is a factor that determines the phase transition of quarks. For the critical temperature, the quark and anti-quark may interact with each other in hadronic phase due to gluon exchange that indicates the strong interaction. The Fermi distribution for quarks systems at hadronic phases is employed to be an analytical evaluation rate of photons. In a quark anti-quark system, the photon rate production at annihilation processes and the transition momentum of the system indicated by the distribution of the quarks, anti-quarks momentum which was governed by the hadronic phase state. The properties of photons yield produced from quark-antiquarks interaction at annihilation processes depend on quarks fugacity and thermal energy of the system.

Hence the effect of fugacity on the photons rate production increase with increases in the fugacity and vice versa. Since we're mainly interested in evaluating the photons rate according to perturbative approximation method that formulated in Eq. (16) that indicates that quarks system are perturbative due to large transverse momentum. The momentum of system equivalent to thermal energy T = 150 MeV, T = 175 MeV, T=200 MeV, T=225 MeV and T=250MeV are P=1.2, 1.4, 1.6, 1.8 and 2.0GeV with critical temperature $T_c$= 155 MeV and 185 MeV for quarks interaction in hadronic phase for the $s\bar{b} \rightarrow \gamma g$ systems.

The photons start emitting from quarks interaction systems until they reach equilibrium at hadronic phase under the time-space. It may be showing that the fugacity coefficient effects on the emission of the photons which result increasingly in annihilation processes it is good to suggest that photons yield production mechanisms could be affected by fugacity and it's producing at quarks interaction media having energy in the limit $4 \leq E_\gamma \leq 8\text{GeV}$. The fugacity effects on the rate of photons are shown in the tables (2) and (3) and figures (1) and (2) correspond to the $s\bar{b} \rightarrow \gamma g$ quarks system at critical temperature 155MeV. Both tables (4) and (5) and figures (3) and (4) show that similar behavior at another critical temperature 185MeV, that the photons rate production enhancement with decreasing critical temperature.

Moreover, it is well-known that photons rate $\mathfrak{F}_{ANN}(\alpha_{ASC}, E_\gamma, T)$ can be evaluated according to the perturbative method as results of collision of flow of quarks. The numerical results of evaluation of the photons rate production are done for the $s\bar{b} \rightarrow \gamma g$ systems. The activation coupling strength constant is evaluated through the different critical temperature and flavor of quarks. So far, we can show from results that the activation of strength coupling and fugacity are the effective quantity in the photons rate calculations.

**Conclusion**

In conclusions, we can find the flow photons emission from quarks interaction is proportional with the electro-color quantum term $\frac{30\ _{em}\alpha_{ASC}}{27\pi^2}$, this refers that photons emission effected by both electrodynamic effect from $\alpha_{em}$ and its effect with color quantum chromodynamics from flavor $N_f$ and color quantum number. The photonic rate through the quark anti-quark interaction at annihilation processes has taken into account the effect of electric charge of quarks system. It increases with increases in electric charge of quarks system. The evaluation is performed for strange anti-bottom interaction quarks for flavor quantum number $N_f$. It means that an increase of active strong coupling leads to a decrease in the photonic rate. It has been found that the flow photons rate proportionally increases with the thermal energy of the system. In case of temperature, the rate is increasing with respect to the decreases of the active coupling while the rate increases with increases to temperature, which indicates that the rate represents functions to temperature. Also, the rate data are founded to proportionally increase with the increases the thermal energy of the system. On the other hand, the rate of photons is affected by the critical temperature and fugacity parameters, as the rate of emission increases by decreasing the critical temperature, while the increases in the fugacity parameters leads to increases in the rate of photons emission.

**Table (1): Result of active of strong coupling for $s\bar{b} \rightarrow \gamma g$, annihilation with critical temperatures $T_c$= 155 MeV and 185 MeV**

| The momentum transition $P_{Em}$ Gev | $\alpha_{ASC}(P_{EM})$ at $T_c =$ 155MeV | $\alpha_{ASC}(P_{EM})$ at $T_c =$ 185MeV |
|---|---|---|
| 1.200 | 0.5417 | 0.5930 |
| 1.400 | 0.5038 | 0.5478 |
| 1.600 | 0.4749 | 0.5139 |
| 1.800 | 0.4521 | 0.4873 |
| 2.000 | 0.4335 | 0.4657 |

**Table (2): Data results of the current photonic rate for the $s\bar{b} \to \gamma g$ system at annihilation processes in critical temperature $T_c = 155 MeV, \lambda_q = 0.02$.**

| $E\gamma$(GeV) | $\mathfrak{F}_{ANN}(\alpha_{ASC}, E_\gamma, T) \frac{1}{GeV^2 fm^4}$ | | | | |
|---|---|---|---|---|---|
| | T=150MeV | T=175 MeV | T=200 MeV | T=225 MeV | T=250 MeV |
| | $P_{EM} = 1.200$ GeV | $P_{EM} = 1.400$ GeV | $P_{EM} = 1.600$ GeV | $P_{EM} = 1.800$ GeV | $P_{EM} = 2.000$ GeV |
| | $\alpha_{ASC} = 0.5417$ | $\alpha_{ASC} = 0.5038$ | $\alpha_{ASC} = 0.4749$ | $\alpha_{ASC} = 0.4521$ | $\alpha_{ASC} = 0.4335$ |
| 4 | 1.12300E-21 | 5.32768E-20 | 9.29381E-19 | 8.15452E-18 | 4.30202E-17 |
| 4.5 | 4.98644E-23 | 3.96139E-21 | 1.03911E-19 | 1.28914E-18 | 9.36866E-18 |
| 5 | 2.09168E-24 | 2.73831E-22 | 1.05578E-20 | 1.79006E-19 | 1.69726E-18 |
| 5.5 | 8.47133E-26 | 1.81332E-23 | 1.01725E-21 | 2.32516E-20 | 2.82262E-19 |
| 6 | 3.35083E-27 | 1.16761E-24 | 9.47872E-23 | 2.90092E-21 | 4.46942E-20 |
| 6.5 | 1.30327E-28 | 7.37250E-26 | 8.63285E-24 | 3.52367E-22 | 6.85721E-21 |
| 7 | 5.00562E-30 | 4.58868E-27 | 7.73397E-25 | 4.19980E-23 | 1.02933E-21 |
| 7.5 | 1.90406E-31 | 2.82492E-28 | 6.84339E-26 | 4.93589E-24 | 1.52069E-22 |
| 8 | 7.18748E-33 | 1.72425E-29 | 5.99746E-27 | 5.73886E-25 | 2.21962E-23 |

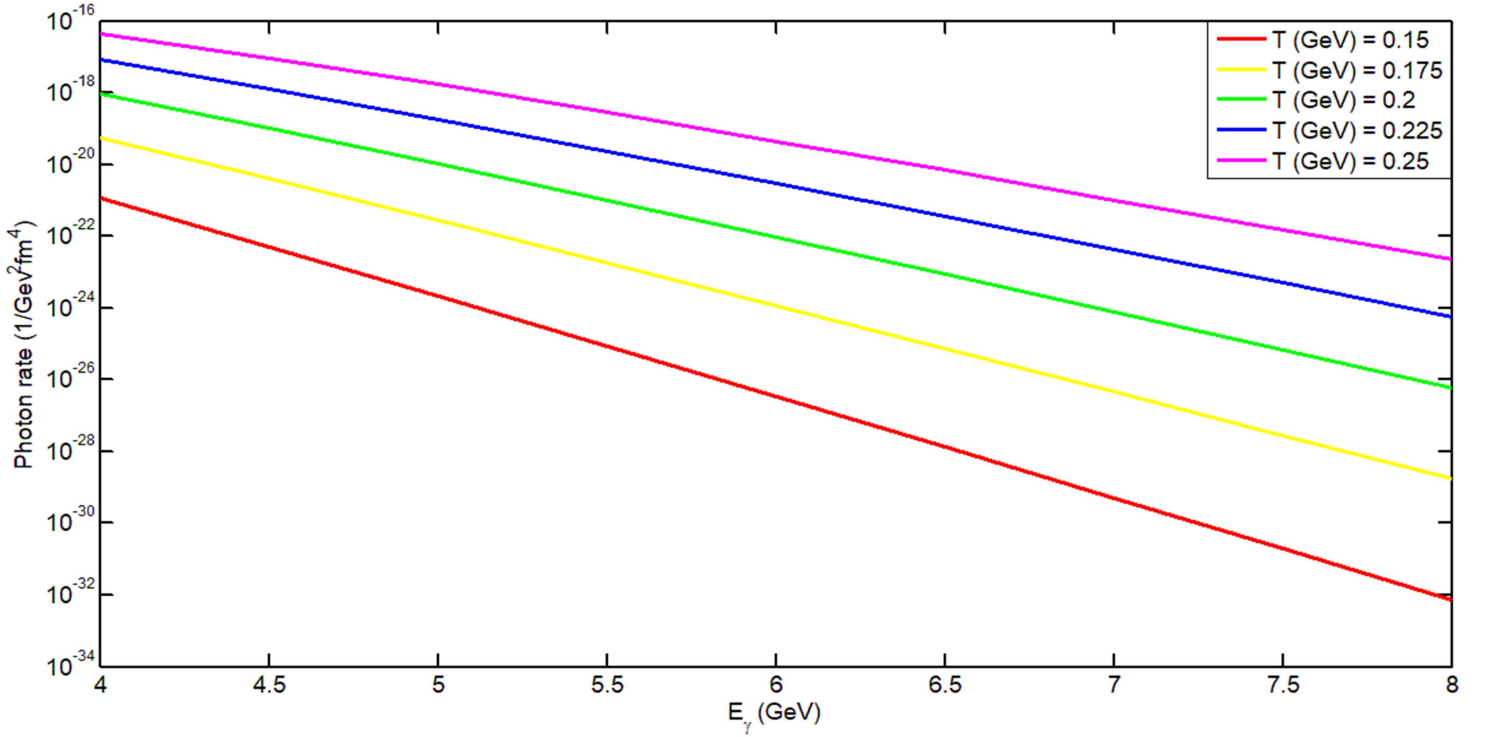

Figure (1): The photon rate $\mathfrak{F}_{ANN}(\alpha_{ASC}, E_\gamma, T)$ as a function of Energy ($E\gamma$) for the system ($s\bar{b} \to \gamma g$) at $T_c$= 155 MeV, $\lambda_q = 0.02$ .

**Table (3): Data results of the current photonic rate for the $s\bar{b} \to \gamma g$ system at annihilation processes in critical temperature $T_c = 155 MeV, \lambda_q = 0.068$.**

| $E\gamma$(GeV) | $\Im_{ANN}(\alpha_{ASC}, E_\gamma, T) \frac{1}{GeV^2 fm^4}$ | | | | |
|---|---|---|---|---|---|
| | T=150MeV | T=175 MeV | T=200 MeV | T=225 MeV | T=250 MeV |
| | $P_{EM} = 1.200$ GeV | $P_{EM} = 1.400$ GeV | $P_{EM} = 1.600$ GeV | $P_{EM} = 1.800$ GeV | $P_{EM} = 2.000$ GeV |
| | $\alpha_{ASC} = 0.5417$ | $\alpha_{ASC} = 0.5038$ | $\alpha_{ASC} = 0.4749$ | $\alpha_{ASC} = 0.4521$ | $\alpha_{ASC} = 0.4335$ |
| 4 | 1.29819E-20 | 6.15880E-19 | 1.07436E-17 | 9.42662E-17 | 4.97313E-16 |
| 4.5 | 5.76432E-22 | 4.57936E-20 | 1.20121E-18 | 1.49025E-17 | 1.08302E-16 |
| 5 | 2.41798E-23 | 3.16548E-21 | 1.22048E-19 | 2.06931E-18 | 1.96203E-17 |
| 5.5 | 9.79286E-25 | 2.09620E-22 | 1.17594E-20 | 2.68788E-19 | 3.26294E-18 |
| 6 | 3.87356E-26 | 1.34976E-23 | 1.09574E-21 | 3.35346E-20 | 5.16665E-19 |
| 6.5 | 1.50658E-27 | 8.52261E-25 | 9.97958E-23 | 4.07336E-21 | 7.92693E-20 |
| 7 | 5.78650E-29 | 5.30451E-26 | 8.94047E-24 | 4.85497E-22 | 1.18990E-20 |
| 7.5 | 2.20109E-30 | 3.26561E-27 | 7.91096E-25 | 5.70589E-23 | 1.75791E-21 |
| 8 | 8.30872E-32 | 1.99323E-28 | 6.93306E-26 | 6.63412E-24 | 2.56588E-22 |

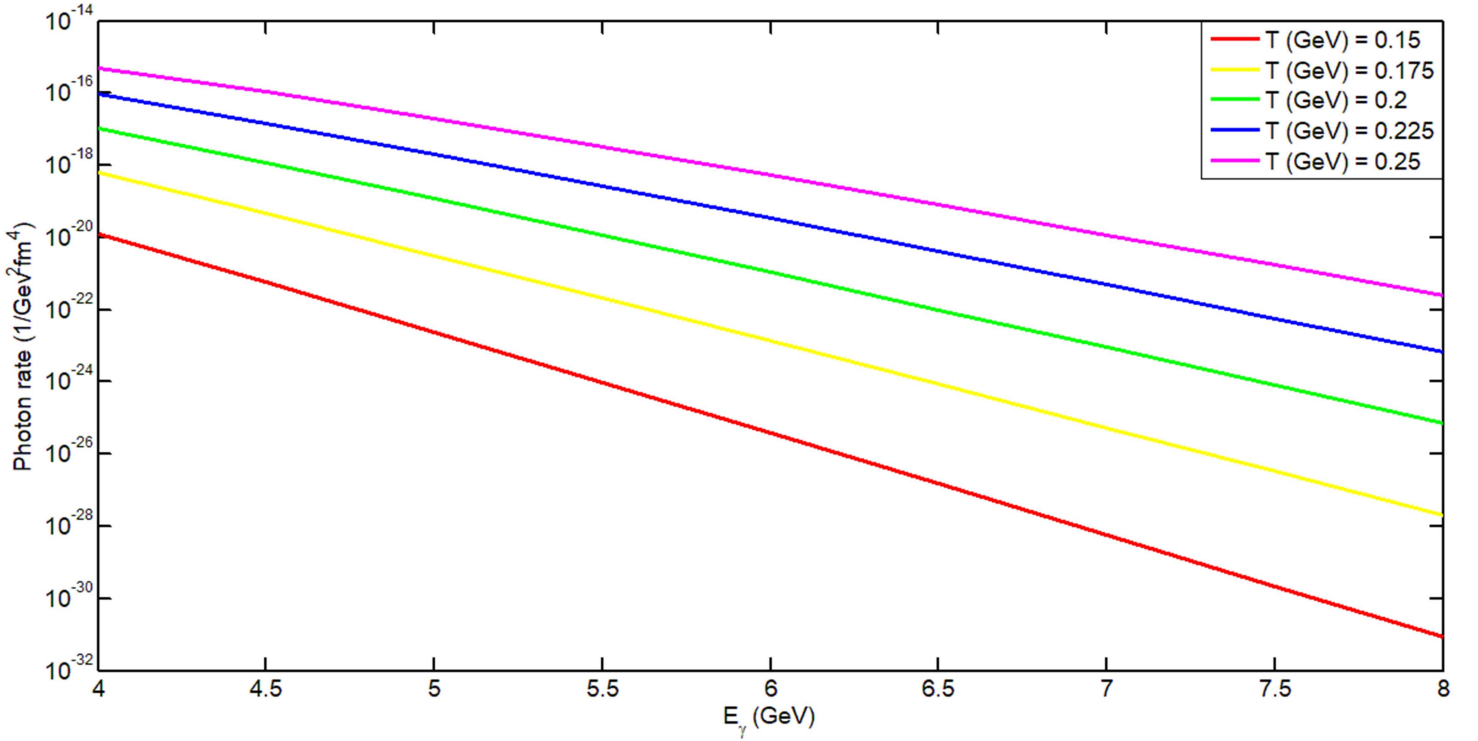

Figure (2): The photon rate $\Im_{ANN}(\alpha_{ASC}, E_\gamma, T)$ as a function of Energy ($E\gamma$) for the system ($s\bar{b} \to \gamma g$) at $T_c$= 155 MeV, $\lambda_q = 0.068$.

**Table (4): Data results of the current photonic rate for the $s\bar{b} \to \gamma g$ system at annihilation processes in critical temperature $T_c = 185 MeV, \lambda_q = 0.02$.**

| $E\gamma$(GeV) | $\mathfrak{F}_{ANN}(\alpha_{ASC}, E_\gamma, T) \frac{1}{GeV^2 fm^4}$ | | | | |
|---|---|---|---|---|---|
| | T=150MeV | T=175 MeV | T=200 MeV | T=225 MeV | T=250 MeV |
| | $P_{EM} = 1.200$ GeV | $P_{EM} = 1.400$ GeV | $P_{EM} = 1.600$ GeV | $P_{EM} = 1.800$ GeV | $P_{EM} = 2.000$ GeV |
| | $\alpha_{ASC} = 0.5930$ | $\alpha_{ASC} = 0.5478$ | $\alpha_{ASC} = 0.5139$ | $\alpha_{ASC} = 0.4873$ | $\alpha_{ASC} = 0.4657$ |
| 4 | 9.98268E-22 | 4.57958E-20 | 7.61692E-19 | 6.22230E-18 | 2.90903E-17 |
| 4.5 | 4.63431E-23 | 3.61047E-21 | 9.24149E-20 | 1.11132E-18 | 7.74689E-18 |
| 5 | 1.99568E-24 | 2.57722E-22 | 9.78072E-21 | 1.62796E-19 | 1.50966E-18 |
| 5.5 | 8.22448E-26 | 1.74182E-23 | 9.65829E-22 | 2.17949E-20 | 2.60778E-19 |
| 6 | 3.29390E-27 | 1.13757E-24 | 9.14933E-23 | 2.77274E-21 | 4.22691E-20 |
| 6.5 | 1.29317E-28 | 7.25816E-26 | 8.43246E-24 | 3.41435E-22 | 6.58907E-21 |
| 7 | 5.00337E-30 | 4.55397E-27 | 7.62267E-25 | 4.11102E-23 | 1.00056E-21 |
| 7.5 | 1.91446E-31 | 2.82154E-28 | 6.79268E-26 | 4.86963E-24 | 1.49124E-22 |
| 8 | 7.26200E-33 | 1.73120E-29 | 5.98704E-27 | 5.69741E-25 | 2.19177E-23 |

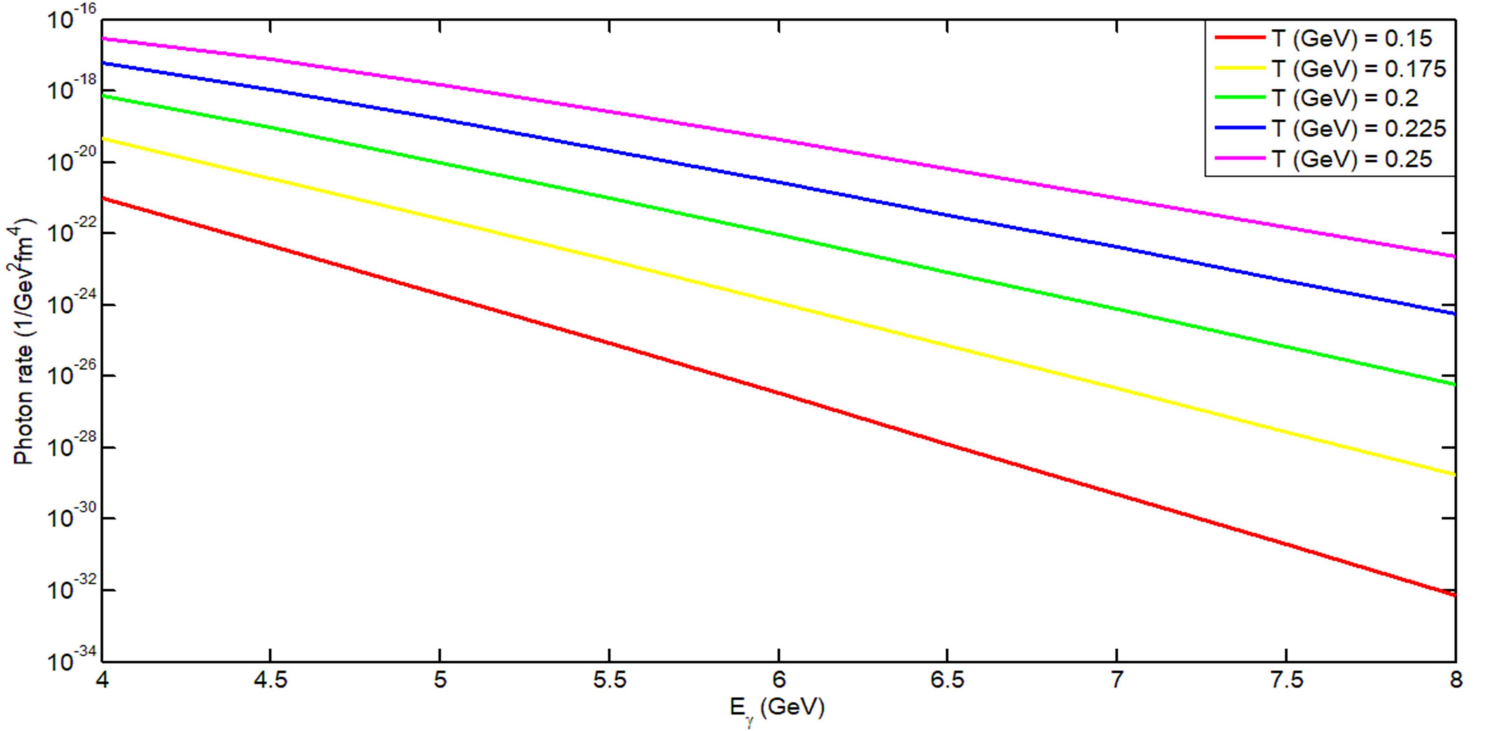

Figure (3): The photon rate $\mathfrak{F}_{ANN}(\alpha_{ASC}, E_\gamma, T)$ as a function of Energy ($E\gamma$) for the system ($s\bar{b} \to \gamma g$) at $T_c$= 185 MeV, $\lambda_q = 0.02$.

**Table (5): Data results of the current photonic rate for the $s\bar{b} \to \gamma g$ system at annihilation processes in critical temperature $T_c = 185 MeV, \lambda_q = 0.068$.**

| $E\gamma$(GeV) | $\mathfrak{F}_{ANN}(\alpha_{ASC}, E_\gamma, T) \frac{1}{GeV^2 fm^4}$ | | | | |
|---|---|---|---|---|---|
| | T=150 MeV | T=175 MeV | T=200 MeV | T=225 MeV | T=250 MeV |
| | $P_{EM} = 1.200$ GeV | $P_{EM} = 1.400$ GeV | $P_{EM} = 1.600$ GeV | $P_{EM} = 1.800$ GeV | $P_{EM} = 2.000$ GeV |
| | $\alpha_{ASC} = 0.5930$ | $\alpha_{ASC} = 0.5478$ | $\alpha_{ASC} = 0.5139$ | $\alpha_{ASC} = 0.4873$ | $\alpha_{ASC} = 0.4657$ |
| 4 | 1.15400E-20 | 5.29400E-19 | 8.80516E-18 | 7.19298E-17 | 3.36284E-16 |
| 4.5 | 5.35726E-22 | 4.17371E-20 | 1.06832E-18 | 1.28469E-17 | 8.95540E-17 |
| 5 | 2.30701E-23 | 2.97927E-21 | 1.13065E-19 | 1.88192E-18 | 1.74517E-17 |
| 5.5 | 9.50750E-25 | 2.01355E-22 | 1.11650E-20 | 2.51949E-19 | 3.01460E-18 |
| 6 | 3.80775E-26 | 1.31503E-23 | 1.05766E-21 | 3.20529E-20 | 4.88631E-19 |
| 6.5 | 1.49491E-27 | 8.39043E-25 | 9.74792E-23 | 3.94699E-21 | 7.61696E-20 |
| 7 | 5.78390E-29 | 5.26439E-26 | 8.81180E-24 | 4.75234E-22 | 1.15665E-20 |
| 7.5 | 2.21312E-30 | 3.26170E-27 | 7.85234E-25 | 5.62930E-23 | 1.72387E-21 |
| 8 | 8.39487E-32 | 2.00127E-28 | 6.92102E-26 | 6.58621E-24 | 2.53368E-22 |

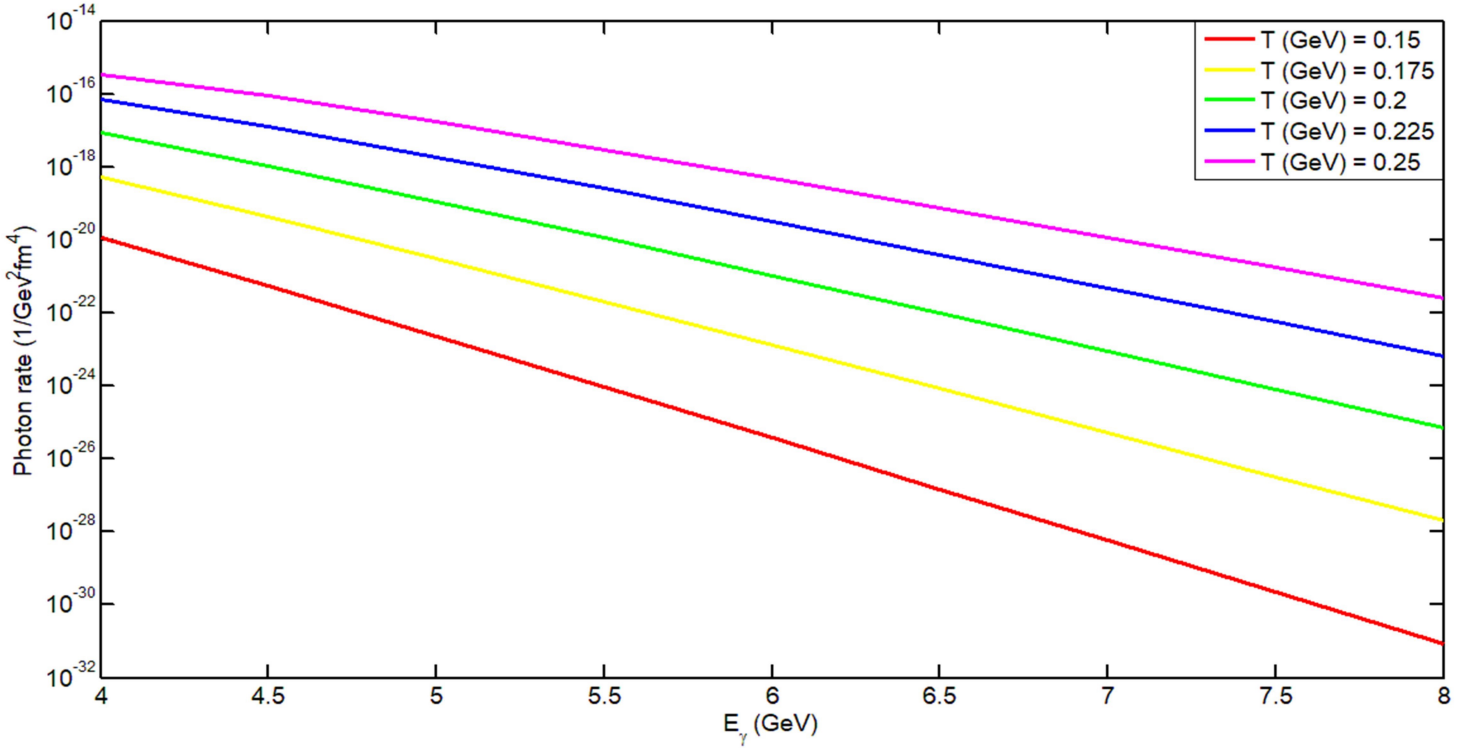

Figure (4): The photon rate $\mathfrak{F}_{ANN}(\alpha_{ASC}, E_\gamma, T)$ as a function of Energy ($E\gamma$) for the system ($s\bar{b} \to \gamma g$) at $T_c$= 185 MeV, $\lambda_q = 0.068$.